# Unique optical characteristics of a Fabry-Perot resonator with embedded *PT*-symmetrical grating

Mykola Kulishov,[1,*] Bernard Kress,[2] and H. F. Jones[3]

[1]*HTA Photomask, 1605 Remuda Lane, San Jose, California 95112, USA*
[2]*Google, 1600 Amphitheatre Parkway, Mountain View, California 94043, USA*
[3]*Physics Department, Imperial College, London, SW7 2BZ, UK*
[*]*mykolak@htaphotomask.com*

**Abstract:** We explore the optical properties of a Fabry-Perot resonator with an embedded Parity-Time (PT) symmetrical grating. This PT-symmetrical grating is non diffractive (transparent) when illuminated from one side and diffracting (Bragg reflection) when illuminated from the other side, thus providing a unidirectional reflective functionality. The incorporated PT-symmetrical grating forms a resonator with two embedded cavities. We analyze the transmission and reflection properties of these new structures through a transfer matrix approach. Depending on the resonator geometry these cavities can interact with different degrees of coherency: fully constructive interaction, partially constructive interaction, partially destructive interaction, and finally their interaction can be completely destructive. A number of very unusual (exotic) nonsymmetrical absorption and amplification behaviors are observed. The proposed structure also exhibits unusual lasing performance. Due to the PT-symmetrical grating, there is no chance of mode hopping; it can lase with only a single longitudinal mode for any distance between the distributed reflectors.

## 1. Introduction

The main types of fiber and solid-state lasers are based on distributed-feedback (DFB) and distributed-Bragg reflector (DBR) architectures. DFB lasers can provide a good performance, especially narrow linewidth and low relative intensity noise (RIN). However, the length of the DFB structure is limited by the Bragg grating manufacturing method. The very short cavity provides the necessary high-frequency separation between longitudinal modes, but it limits the absorption efficiency of pump power, which results in the output power being limited. To improve the output power and efficiency of the laser, the cavity length needs to be increased. A DBR laser can increase its cavity length to get a higher gain. However, when the cavity length of the DBR exceeds a certain value, multiple longitudinal modes appear in the reflection spectrum of distributed Bragg reflectors.

Traditionally the gain is distributed uniformly in between the reflectors. However, it has been shown recently that spatially inhomogeneous gain inside the resonator cavity results in the existence of modes that do not lase near passive cavity resonance frequencies and do not evolve smoothly out of passive cavity modes [1]. Even more exotic properties can be produced when the cavity is divided not between gain and gain-free areas, but between gain and loss accompanied by index modulation such that parity-time (*PT*) symmetry is preserved. In optics, the *PT* symmetry concept, which originated from quantum physics, requires that the complex refractive index of the structures satisfy the following equation: $n(z) = n^*(-z)$ [3], where (*) stands for complex conjugation. Such optical structures with balanced gain and loss were the subject of intense study even before *PT*-symmetry concept was introduced by C.M. Bender et al. in 1999 [3] and long before this concept was introduced into optics. In fact, L. Poladian was the first to present this concept for transverse geometry $n(z) = n^*(-z)$ reflective Bragg gratings, in 1996 [4], and in the same year H.-P. Nolting et al. described waveguides with $n(x, y) = n^*(-x, -y)$ longitudinal geometry [5]. Later this concept was studied in detail for contradirectional [6] and codirectional [7] mode coupling. *PT*-symmetrical reflective gratings with balanced gain and loss, $\Delta n(z) = a\cos(kz) + jb\sin(kz)$, near the symmetry-breaking point ($a = b$) do not reflect input light launched from one side, exhibiting complete invisibility [6] and making it impossible to form a cavity. However, when two such *PT*-symmetric gratings are concatenated, with the reflecting ends separated by a non-modulated portion of the waveguide, their unidirectional behavior can be used to build DFB/DBR structures that differ significantly from those traditionally used, with unique characteristics including inherently single lasing mode operation [8]. On the other hand, a concatenation of a PT-symmetric Bragg grating and a traditional Bragg grating produces a series of exotic nonreciprocal absorption and amplification behaviors [9] that makes these combined structures more intriguing and practical for building optical sensors, coherent perfect absorbers (CPA) [10], thermal detectors etc.

As a next logical step, in this paper we study the Distributed Bragg Reflector (DBR) formed by two traditional Bragg gratings where a traditional uniform gain medium in the cavity is replaced by the active periodical structure of a PT-symmetrical Bragg grating (PT-SBG) in its balanced form with zero net single pass gain.

## 2. Transfer Matrix for Fabry-Perot cavity with a single *PT*-SBG

We consider the FP resonator shown in Fig.1, where a cavity is formed by two (left and right) traditional Bragg gratings with length $L_1$ and $L_2$, and coupling coefficients $\kappa_1$ and $\kappa_2$ respectively. A PT-SBG with length $L_{PT}$ and coupling coefficient $\kappa_{PT}$ is placed inside the cavity with distances $d_1$ and $d_2$ between the left and right gratings. The PT-symmetric Bragg grating is formed by the combined periodical modulation of refractive index and gain/loss with a quarter-period shift between them:

$$\Delta n(z) = a\cos(kz) + jb\sin(kz) \qquad (1)$$

where $k = 2\pi/\Lambda$, and the $z$ is the coordinate along the propagation direction, and all three gratings have the same period of index modulation $\Lambda$. At the symmetry-breaking point when the two amplitudes are equal ($a = b$) this *PT*-SBG exhibits perfect unidirectionality, i.e. an input wave launched from the left end of the grating is transmitted completely without any reflection or change in phase (full invisibility [6]). At the same time this wave launched from the right end exhibits enhanced reflection at the resonance if $\kappa_{PT}L_{PT} > 1$ [6], however, the transmitted portion again does not change its amplitude or phase. The transmission characteristics of the *PT*-SBG can be described by the following transfer matrix [6]:

$$M_{PT} = \begin{bmatrix} \exp(j\beta L_{PT}) & m_{12}^{PT} \exp(-j(\sigma-\beta)L_{PT}) \\ 0 & \exp(-j\beta L_{PT}) \end{bmatrix}, \qquad (2)$$

where $\beta$ is the propagation constant of the propagated light wave, and $\sigma = \beta - \pi/\Lambda$ is the phase mismatch factor and

$$m_{12}^{PT} = j(\kappa_{PT}/\sigma)\sin(\sigma L_{PT}), \qquad (3)$$

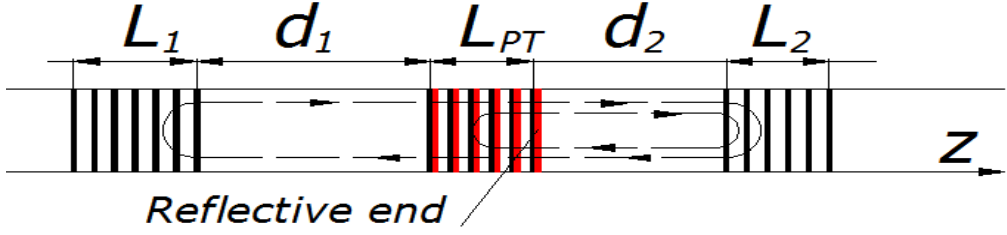

Fig. 2. Geometry for a FP structure formed by two traditional Bragg gratings, $L_1$ and $L_2$, with a $PT$-SBG, $L_{PT}$, placed inside the cavity.

At the same time the transfer matrix of the traditional (only index modulated) Bragg gratings is described by the following matrix ($i = 1, 2$):

$$M_i = \begin{bmatrix} m_{11}^{(i)} \exp(-j(\sigma-\beta)L_i) & m_{12}^{(i)} \exp(-j(\sigma-\beta)(L_i+2z_i)) \\ m_{21}^{(i)} \exp(j(\sigma-\beta)(L_i+2z_i)) & m_{22}^{(i)} \exp(j(\sigma-\beta)L_i) \end{bmatrix} \qquad (4)$$

where

$$m_{11}^{(i)} = \cosh(\gamma_i L_i) + j\frac{\sigma}{\gamma_i}\sinh(\gamma_i L_i); \qquad m_{12}^{(i)} = j\frac{\kappa_i}{\gamma_i}\sinh(\gamma_i L_i);$$
$$m_{21}^{(i)} = -j\frac{\kappa_i}{\gamma_i}\sinh(\gamma_i L_i); \qquad m_{22}^{(i)} = \cosh(\gamma_i L_i) - j\frac{\sigma}{\gamma_i}\sinh(\gamma_i L_i); \qquad (5)$$

and $\gamma_i = \left(\kappa_i^2 - \sigma^2\right)^{1/2}$. The coordinates $z_i$ of the grating origins appear in the phase factors of the off-diagonal elements. It is easy to see that $z_1 = 0$ and $z_2 = L_1 + d_1 + L_{PT} + d_2$. We neglect waveguide dispersion for this simulation and will use effective refractive index $n_{eff} = 1.55$. If the index modulation of the first (left) grating is described by $\cos(2\pi z/\Lambda)$, then to keep the right grating in phase its index modulation should be presented as $\cos(2\pi(z-t)/\Lambda)$, where $t$ represents the distance between the gratings ($t = d_1 + L_{PT} + d_2$ in the case of the right grating). Such index modulation results in different expressions for the coupling coefficients: $\kappa_1 = \kappa$ and $\kappa_2 = \kappa \exp(\mp 2\pi j t/\Lambda)$. A similar argument applies to the PT grating. Taking into account all these factors, the transfer matrix of the entire FP resonator in Fig.1 can be obtained by matrix multiplication:

$$M_\Sigma = M_2 I(d_2) M_{PT} I(d_1) M_1, \qquad (6)$$

where $I(d_i)$ is the transfer matrix of the unperturbed section of the cavity ($i = 1, 2$):

$$I_i = \begin{bmatrix} \exp(j\beta d_i) & 0 \\ 0 & \exp(-j\beta d_i) \end{bmatrix} \qquad (7)$$

and has the following elements:

$$m_{11}^\Sigma = \left[ m_{11}^{(1)} m_{11}^{(2)} \exp(2j\beta(d_1+d_2+L_{PT})) + m_{21}^{(1)} m_{11}^{(2)} m_{12}^{(PT)} \exp(2j\beta(d_2+L_{PT}))\exp(-j\sigma L_{PT}) + \right.$$
$$\left. + m_{21}^{(1)} m_{12}^{(2)} \right] \exp(j\Phi) \qquad (8a)$$

$$m_{12}^{\Sigma} = \left[ m_{11}^{(2)} m_{12}^{(1)} \exp(2j\beta(d_1 + d_2 + L_{PT})) + m_{11}^{(2)} m_{22}^{(1)} m_{12}^{(PT)} \exp(2j\beta(d_2 + L_{PT})) \exp(-j\sigma L_{PT}) + \right.$$
$$\left. + m_{12}^{(2)} m_{22}^{(1)} \right] \exp(j\Phi) \qquad (8b)$$

$$m_{21}^{\Sigma} = \left[ m_{21}^{(2)} m_{11}^{(1)} \exp(2j\beta(d_1 + d_2 + L_{PT})) + m_{21}^{(1)} m_{21}^{(2)} m_{12}^{(PT)} \exp(2j\beta(d_2 + L_{PT}) \exp(-j\sigma L_{PT}) + \right.$$
$$\left. + m_{21}^{(1)} m_{22}^{(2)} \right] \exp(j\Phi) \qquad (8c)$$

$$m_{22}^{\Sigma} = \left[ m_{21}^{(2)} m_{12}^{(1)} \exp(2j\beta(d_1 + d_2 + L_{PT})) + m_{21}^{(2)} m_{22}^{(1)} m_{12}^{(PT)} \exp(2j\beta(d_2 + L_{PT})) \exp(-j\sigma L_{PT}) + \right.$$
$$\left. + m_{22}^{(1)} m_{22}^{(2)} \right] \exp(j\Phi), \qquad (8d)$$

where $\Phi = -\beta(d_1 + d_2 + L_{PT}) + \pi(L_1 + L_2)/\Lambda$.

It is straightforward to show that the transmission and reflection coefficients for launching a wave from the left side can be expressed as

$$T_{\Sigma}^{(L)} = \left| t_{\Sigma}^{(L)} \right|^2 = \left| \frac{m_{11}^{\Sigma} m_{22}^{\Sigma} - m_{12}^{\Sigma} m_{21}^{\Sigma}}{m_{22}^{\Sigma}} \right|^2; \qquad R_{\Sigma}^{(L)} = \left| r_{\Sigma}^{(L)} \right|^2 = \left| -\frac{m_{21}^{\Sigma}}{m_{22}^{\Sigma}} \right|^2 \qquad (9)$$

and for launching a wave from the right side

$$T_{\Sigma}^{(R)} = \left| t_{\Sigma}^{(R)} \right|^2 = \left| \frac{1}{m_{22}^{\Sigma}} \right|^2; \qquad R_{\Sigma}^{(R)} = \left| r_{\Sigma}^{(R)} \right|^2 = \left| \frac{m_{12}^{\Sigma}}{m_{22}^{\Sigma}} \right|^2 \qquad (10)$$

Actually, it is easy to see that in Eq.(9) the numerator $m_{11}^{\Sigma} m_{22}^{\Sigma} - m_{12}^{\Sigma} m_{21}^{\Sigma} = 1$, since each of the constituent matrices in Eq.(6) has unit determinant, so that the transmission from the left and right side is identical ($T_{\Sigma}^{(R)} = T_{\Sigma}^{(L)} = T_{\Sigma}$). Another important feature of the obtained result is that when the set-up is PT-symmetric it satisfies the modified unitarity condition [11]:

$$|T_{\Sigma} - 1| = \sqrt{R_{\Sigma}^{(R)} R_{\Sigma}^{(L)}} \qquad (11)$$

This forms an important check of our calculations.

For a laser oscillator without injected signal the lasing condition is defined by $m_{22}^{\Sigma}(\lambda) = 0$, which can be presented as

$$\frac{m_{21}^{(2)}}{m_{22}^{(2)}} \left[ \frac{m_{12}^{(1)}}{m_{22}^{(1)}} \exp(2j\beta(d_1 + d_2 + L_{PT})) + m_{12}^{(PT)} \exp(2j\beta(d_2 + L_{PT})) \exp(-j\sigma L_{PT}) \right] + 1 = 0 \qquad (12)$$

This threshold condition can be expressed in more traditional form through the reflection coefficients of the regular Bragg gratings (left and right) and the central PT-SBG:

$$r_2^{(L)} \left( r_1^{(R)} \exp(2j\beta(d_1 + d_2 + L_{PT}) + r_{PT}^{(R)} \exp(2j\beta(d_2 + L_{PT})) \right) = 1 \qquad (13)$$

where $r_2^{(L)} = \left| m_{21}^{(2)} / m_{22}^{(2)} \right|$, $r_1^{(R)} = \left| m_{12}^{(1)} / m_{22}^{(1)} \right|$ and $r_{PT}^{(R)} = \left| m_{12}^{(PT)} \right|$.

In the case of zero index and gain/loss modulation between the left and right Bragg reflectors, $\kappa_{PT} = 0$, and Eq.(13) reduces to the traditional equation for a DBR structure:

$$r_1^{(R)} r_2^{(L)} \exp(2j\beta(d_1 + d_2 + L_A)) = 1, \qquad (14)$$

where lasing is achieved through the introduction of the gain in the waveguide between the Bragg reflectors: $\beta = \beta' + j\alpha$ with $\alpha < 0$.

On the other hand, Eq.(13) includes the case of the cavity between the left Bragg reflector and the PT-SBG without the right Bragg reflector, $r_1^{(R)} = 0$:

$$r_2^{(L)} r_{PT}^{(R)} \exp(2j\beta(d_2 + L_{PT})) = 1 \qquad (15)$$

This cavity can generate lasing with zero average gain/loss level when $\kappa_{PT}L_{PT} \geq 1$. Overall Eq.(13) describes the threshold condition for two embedded cavities created by the PT-SBG placed between the traditional Bragg reflectors.

These two embedded cavities can interact differently depending on the geometrical parameters of the DBR structure. With zero average gain in the cavity the two phase factors in (13) are unimodular and can take certain values depending on the length of $d_1$, $d_2$ and $L_{PT}$. There are four simple cases, (a) - (c), where the phase factors are equal to $\pm 1$, with some possible choices of the lengths $d_1$, $d_2$ and $L_{PT}$:

$$\left.\begin{aligned} \exp(2j\beta(d_1+d_2+L_{PT})) &= -1 \\ \exp(2j\beta(d_2+L_{PT})) &= -1 \end{aligned}\right\} d_1 = m\Lambda; d_2 = (p+1/2)\Lambda; L_{PT} = q\Lambda \quad (16a)$$

$$\left.\begin{aligned} \exp(2j\beta(d_1+d_2+L_{PT})) &= 1 \\ \exp(2j\beta(d_2+L_{PT})) &= -1 \end{aligned}\right\} d_1 = (m+1/2)\Lambda; d_2 = (p+1/2)\Lambda; L_{PT} = q\Lambda \quad (16b)$$

$$\left.\begin{aligned} \exp(2j\beta(d_1+d_2+L_{PT})) &= -1 \\ \exp(2j\beta(d_2+L_{PT})) &= 1 \end{aligned}\right\} d_1 = (m+1/2)\Lambda; d_2 = p\Lambda; L_{PT} = q\Lambda \quad (16c)$$

$$\left.\begin{aligned} \exp(2j\beta(d_1+d_2+L_{PT})) &= 1 \\ \exp(2j\beta(d_2+L_{PT})) &= 1 \end{aligned}\right\} d_1 = m\Lambda; d_2 = p\Lambda; L_{PT} = q\Lambda \quad (16d)$$

where *m, p, q* are integers. Eqs.(16) are a set of different phase conditions that make the cavities interact constructively or destructively. The first two phase arrangements, (16a) and (16b), are capable of producing a lasing effect. The first, (16a), is characterized by a very low threshold condition, $\kappa_{PT}L_{PT} < 1$, or in other words with a rather weak PT-SBG that does not generate any amplification by itself (standing alone); we can explain it by a constructive interaction between the two cavities within the FP structure. The second condition, (16b), is capable of producing lasing, i.e. $m_{22}^{\Sigma}(\lambda) = 0$, but with a very strong PT-SBG, $\kappa_{PT}L_{PT} > 1$, i.e. when this grating produces amplification operating independently. We will call this condition partially constructive cavity interaction. The next condition, (16c), does not provide any solution to $m_{22}^{\Sigma}(\lambda) = 0$, and is therefore incapable of producing any lasing effect. However, in this arrangement it is possible to achieve full light absorption at the resonance wavelength, as we will show below. Therefore, we will term this condition "partially destructive" cavity interaction. Finally, condition (16d) can be characterized as fully destructive cavity interaction, and the FP structure in this condition does not produce any lasing or other interesting optical behavior.

### 3. Absorption and Amplification modes along with lasing characteristics

*3.1 Fully constructive cavity interaction*

The condition (16a) provides fully constructive interference of the optical field from the two cavities with right-side reflection, left-side reflection and transmission coefficients at the resonance wavelength ($\lambda_R = 1550$ nm, $\sigma = 0$) that can be written in the following form:

$$\left|r_{PT}^{(left)}\right| = \left|\frac{\tanh(\kappa_2 L_2) + \kappa_{PT} L_{PT} \tanh(\kappa_1 L_1)\tanh(\kappa_2 L_2) - \tanh(\kappa_1 L_1)}{1 - \tanh(\kappa_1 L_1)\tanh(\kappa_2 L_2) - \kappa_{PT} L_{PT} \tanh(\kappa_2 L_2)}\right|;$$

$$\left|r_{PT}^{(right)}\right| = \left|\frac{\tanh(\kappa_1 L_1) + \kappa_{PT} L_{PT} - \tanh(\kappa_2 L_2)}{1 - \tanh(\kappa_1 L_1)\tanh(\kappa_2 L_2) - \kappa_{PT} L_{PT} \tanh(\kappa_2 L_2)}\right|; \qquad (17)$$

$$\left|t_{PT}\right| = \left|\frac{\text{sech}(\kappa_1 L_1)\text{sech}(\kappa_2 L_2)}{1 - \tanh(\kappa_1 L_1)\tanh(\kappa_2 L_2) - \kappa_{PT} L_{PT} \tanh(\kappa_2 L_2)}\right|.$$

*3.1.1 Lasing*

Lasing is achieved when

$$\kappa_{PT} L_{PT} = \coth(\kappa_2 L_2) - \tanh(\kappa_1 L_1) \qquad (18)$$

A comparison of the transmission and reflection spectra of the proposed DBR FP resonator with the PT-SBG inside the cavity with a traditional DBR FP resonator with uniform gain in the cavity ($d_1+L_{PT}+d_2$) between the gratings is presented in Fig.2. The difference is very significant. However, such a behavior has a clear explanation. Unlike, the traditional DBR where uniformly distributed gain potentially amplifies a signal in a broad spectral range, the amplifying bandwidth of the PT-SBG can be much narrower. Indeed, as long as the bandwidth of the PT-SBG is within the free spectral range of the resonator the threshold condition (18) can be satisfied only at a single lasing mode without any possibility of mode hopping. Such a behavior can be called endlessly single-mode lasing.

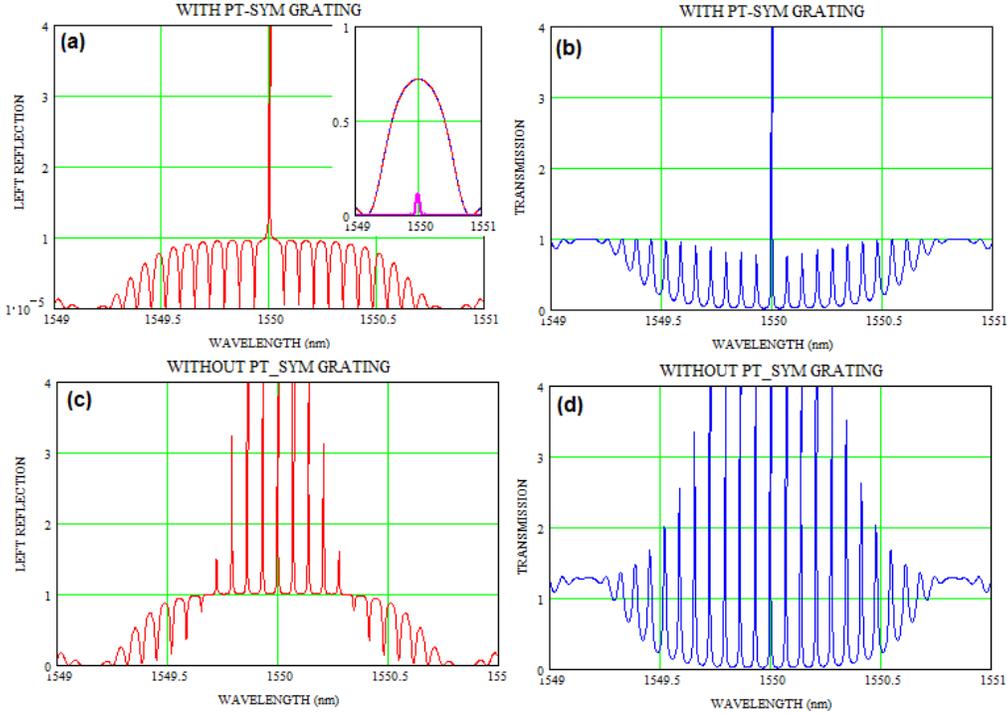

Fig. 2. Fully constructive interference: left reflection (a, c) and transmission (b, d) spectra of the DBG FP resonator with the PT-SBG grating inside the cavity and of the DBR FP resonator without PTSB (b, d) and with an amplifying waveguide ($\alpha = -12$ μm$^{-1}$). The structure parameters are: $\Lambda = 0.5$ μm; $L_1 = L_2 = 2000\Lambda = 1$ mm; $\kappa_1 = \kappa_2 = 1257$ m$^{-1}$; $L_{PT} = 20000\Lambda = 10000$ μm; $\kappa_{PT} = 32.67$ m$^{-1}$; $d_1 = 100\Lambda = 50$ μm; $d_2 = 1000.5\Lambda = 500.25$ μm; The insert in (a) shows the reflection spectra of the left (blue), right (red) and PT-symmetric Bragg grating (magenta).

In traditional lasers the existing lasing mode evolves into the fundamental lasing mode and subsequently into higher order lasing modes that are determined by the geometrical parameters of the passive resonator, the so called "cold cavity", and its permittivity in the absence of gain. These modes are selected to lase due to their large spatial and spectral overlap when gain is placed into the resonator cavity. The recently proposed approach to add into a traditional FP cavity a filtering element in form of with a π-phase shifted FBG [12] to pass only a single lasing mode and block excitation of other longitudinal modes is a partial solution. The proposed DBR with the PT-SBG offers a distributed gain structure that produces very different mode lasing behavior, and a very different lasing threshold condition. In traditional DBR structures any laser media have a relatively wide spectrum of amplification (tens of nm). On the other hand, the proposed PT-SBG design allows one to make the amplifying spectral bandwidth extremely narrow, which in turn leads to very low relative intensity noise, and has the potential to act as an enhanced transmission and reflection ultra-narrow filter or amplifier.

If the PT-SBG bandwidth is broader than the cavity free spectral range (for example, a short PT-SBG), all modes with the bandwidth will be amplified and the structure might support more than one lasing mode.

Actually it does not make much sense to talk about transmission or reflection spectra in the case of lasing, because we do not launch any external signal into the structure. However, to suppress any longitudinal modes besides the lasing one in lasing from the right side in Fig.2, the reflection from the right Bragg reflector, $\kappa_2 L_2$, (Figs.3a and 3b) should be close to 100%. Similarly, to suppress other longitudinal modes in lasing from the left side, the reflection from the left Bragg reflector, $\kappa_1 L_1$ (Figs.3c and 3d) should be close to 100%. Obviously the threshold conditions for the two situations shown in Fig.3(a, b) and Fig.3(c, d) are different: $\kappa_{PT} L_{PT} = 1 - \tanh(\kappa_1 L_1)$ and $\kappa_{PT} L_{PT} = \coth(\kappa_2 L_2) - 1$, respectively. As a result, the $\kappa_{PT}$ values are different in these two cases. Whereas the transmission coefficient is practically identical in both configurations (Fig.3b and 3d), the (left) reflection coefficients are quite different, with much narrower lasing bandwidth in the case of 100% reflection of the left Bragg reflector.

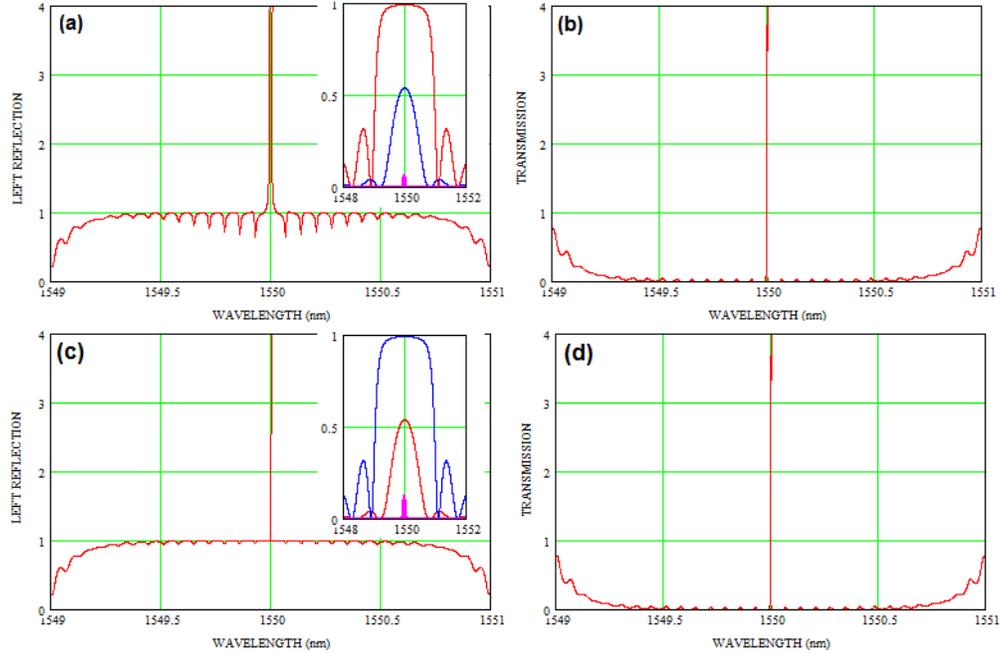

Fig.3. Fully constructive interference: pre-lasing spectra of the DBG FP resonator with the PT-SBG grating inside the cavity. The spacings $\Lambda$=0.5μm, $L_1$ = $L_2$ =2000$\Lambda$ = 1 mm, $L_{PT}$ =20000$\Lambda$=*10000* μm, $d_1$ = 100$\Lambda$ = 50 μm, $d_2$ =1000.5$\Lambda$=*500.25* μm, are common to all panels. In the top panels, with parameters $\kappa_1$ = 942.5 m$^{-1}$, $\kappa_2$=3142 m$^{-1}$, $\kappa_{PT}$ =26.74 m$^{-1}$, reflection from the right Bragg reflector is close to 100% at the resonance wavelength. In the bottom panels, with parameters $\kappa_1$ = 3142 m$^{-1}$, $\kappa_2$=*942.5* m$^{-1}$, $\kappa_{PT}$ =36.12 m$^{-1}$, reflection from the left Bragg reflector is close to 100% at resonance. The Inserts in (a) and (c) show the reflection spectra of the left (blue), right (red) and PT-symmetric Bragg grating (magenta).

*3.1.2 Suppression of left reflection*

The reflection of the DBR structure with the PT-SBG from the left side can be completely suppressed if the PT-SBG strength satisfies the following formula:

$$(\kappa_{PT} L_{PT})_0 = \coth(\kappa_2 L_2) - \coth(\kappa_1 L_1) , \qquad (19)$$

which requires $\kappa_2 L_2 < \kappa_1 L_1$. In that case the transmission and reflection from the right side takes the following form:

$$\left|r_{PT}^{(right)}\right| = \frac{\sinh(\kappa_1 L_1 - \kappa_2 L_2)\cosh(\kappa_1 L_1 + \kappa_2 L_2)}{\sinh^2(\kappa_2 L_2)}; \qquad \left|t_{PT}\right| = \frac{\sinh(\kappa_1 L_1)}{\sinh(\kappa_2 L_2)}. \qquad (20)$$

*3.1.3 Suppression of right reflection*

On the other hand, as follows from Eq.(17), when

$$\kappa_{PT} L_{PT} = \tanh(\kappa_2 L_2) - \tanh(\kappa_1 L_1), \qquad (21)$$

which requires $\kappa_2 L_2 > \kappa_1 L_1$, there is no reflection from the right side at the resonance wavelength, and the transmission and reflection from the left side take the following expressions:

$$\left|r_{PT}^{(left)}\right| = \frac{\sinh(\kappa_2 L_2 - \kappa_1 L_1)\cosh(\kappa_1 L_1 + \kappa_2 L_2)}{\cosh^2(\kappa_1 L_1)}; \qquad \left|t_{PT}\right| = \frac{\cosh(\kappa_2 L_2)}{\cosh(\kappa_1 L_1)}; \qquad (22)$$

Typical spectra for transmission and reflection from the left and right side are shown in Fig.4. It demonstrates very non-symmetrical behavior of the FP resonator with the PT-SBG. Whereas transmitted light goes through the resonator with narrowband amplification, the reflections from the left and right side are completely different. The presence of the PT-SBG with the proper grating strength opens a very narrow transmission window in the left-side reflection (Figs.4(a), (b)) or the right side-reflection (Figs.4(f),(e)). The amplification is achieved with a rather weak PT-SBG with a reflection level just less than 5%, $(\kappa_{PT} L_{PT})^2 \approx 0.03$. Actually, approaching 100% reflectivity ($\tanh(\kappa_1 L_1) \to 1$) of the left reflector, right-side amplification can be extremely high (when the denominator in Eq.(20) approaches zero). Symmetrically, when the right reflector is very strong ($\tanh(\kappa_2 L_2) \to 1$), then we might have extremely amplified narrowband reflection from the left side with zero reflectivity from the right side.

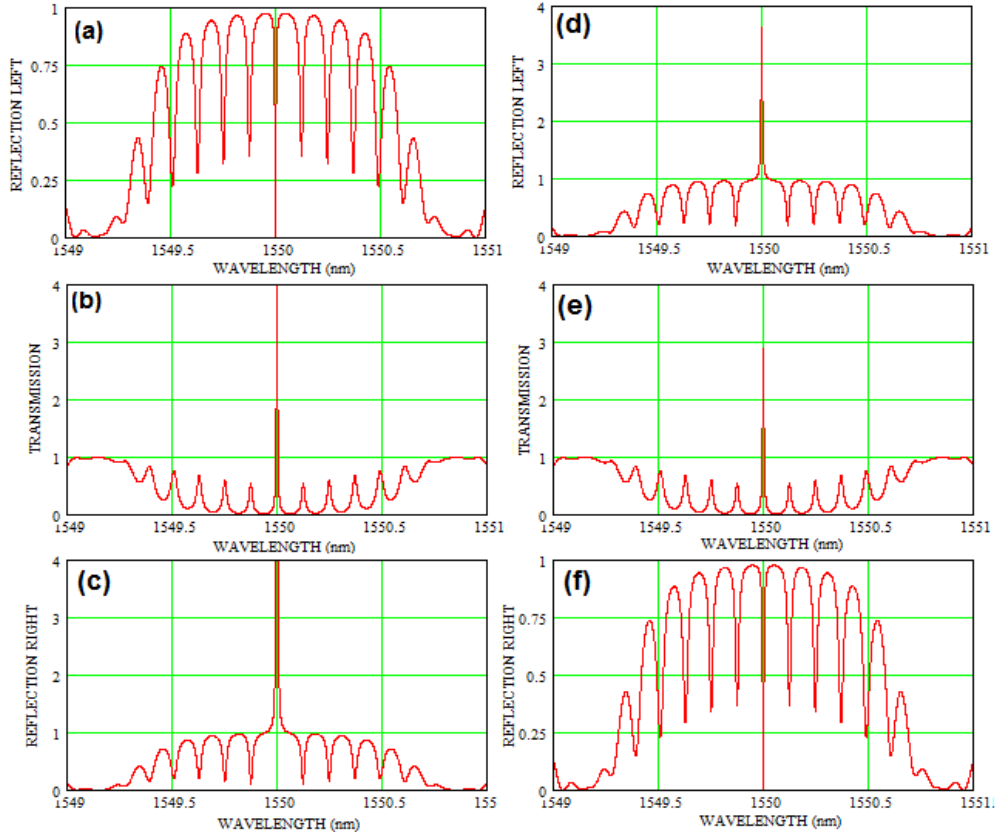

Fig.4. Fully constructive interference: reflection from the left (a,d) and right (c, f) and transmission (b,e) spectra of the FP resonator with the PT-SBG grating inside the cavity The spacing $\Lambda = 0.5$ μm, $L_1 = L_2 = 2000\Lambda = 1$ mm, $L_{PT} = 10000\Lambda = 5000$ μm, $d_1 = 100\Lambda = 50$ μm, $d_2 = 1000.5\Lambda = 500.25$ μm, are common to all panels. For the left panels, with parameters $\kappa_1 = 1571$ m$^{-1}$; $\kappa_2 = 942.5$ m$^{-1}$; $\kappa_{PT} = 53.54$ m$^{-1}$, reflection from the left is completely suppressed at the resonance wavelength, while for the right panels, with parameters $\kappa_1 = 942.5$ m$^{-1}$, $\kappa_2 = 1571$ m$^{-1}$, $\kappa_{PT} = 36$ m$^{-1}$, reflection from the right is completely suppressed at resonance.

*3.2 Partially constructive cavity interaction*

For the partially constructive cavity interaction (Eq.16(b)), the reflection and transmission coefficients at resonance (σ=0) are described by the following expressions:

$$\left|r_{PT}^{(left)}\right| = \left|\frac{\tanh(\kappa_2 L_2) - \kappa_{PT} L_{PT} \tanh(\kappa_1 L_1)\tanh(\kappa_2 L_2) + \tanh(\kappa_1 L_1)}{1 - \kappa_{PT} L_{PT} \tanh(\kappa_2 L_2) + \tanh(\kappa_1 L_1)\tanh(\kappa_2 L_2)}\right|$$

$$\left|r_{PT}^{(right)}\right| = \left|\frac{\tanh(\kappa_1 L_1) - \kappa_{PT} L_{PT} + \tanh(\kappa_2 L_2)}{1 - \kappa_{PT} L_{PT} \tanh(\kappa_2 L_2) + \tanh(\kappa_1 L_1)\tanh(\kappa_2 L_2)}\right|$$

$$\left|t_{PT}\right| = \left|\frac{\text{sech}(\kappa_1 L_1)\text{sech}(\kappa_2 L_2)}{1 - \kappa_{PT} L_{PT} \tanh(\kappa_2 L_2) + \tanh(\kappa_1 L_1)\tanh(\kappa_2 L_2)}\right| \quad (23)$$

### 3.2.1 Lasing

In this case the proposed structure can also operate in a lasing mode when the following equation is satisfied:

$$\kappa_{PT} L_{PT} = \coth(\kappa_2 L_2) + \tanh(\kappa_1 L_1) \qquad (24)$$

This threshold condition requires a much stronger PT-SBG, as can be seen in Fig.5, where the (left) reflection and transmission spectra are shown for different distributed reflector parameters. In the case of Fig5a and Fig.5b the distributed reflectors have equal reflectivity of 73%. As we see in the inset to Fig.5a, unlike the case of fully constructive cavity interaction, the PT-SBG needs to be much stronger and provide significant amplification. It is obvious from Eq.(24) that the lowest threshold value for the partially constructive cavity interaction is to make the right distributed reflector close to 100% reflective, i.e. $\tanh^2(\kappa_2 L_2) \approx 1$, and effectively remove the left reflector, $\kappa_1 L_1 = 0$. The pre-lasing spectra for such a configuration are shown in Fig.5c and 5d for left reflection and transmission respectively. This case of the partially constructive cavity interaction actually requires a much stronger PT-SBG, with a strength that provides close to 4-times amplification at the resonance, as can be seen from the insert in Fig.5a. Its strength can be reduced to an amplification slightly higher than unity when the left Bragg reflector is removed.

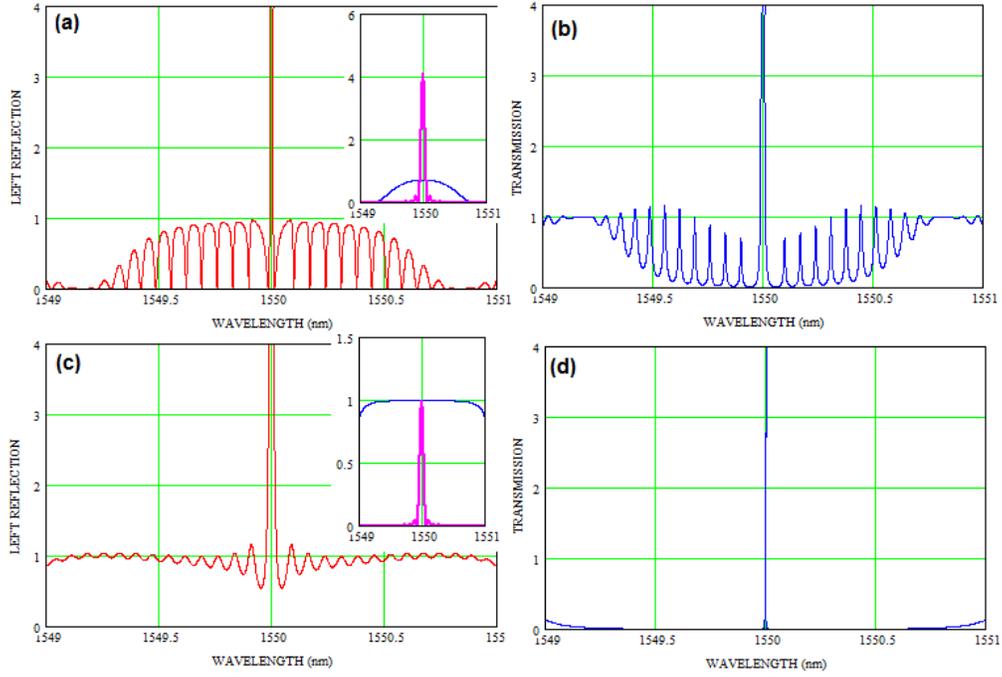

Fig.5. Partially constructive interference: pre-lasing spectra of the DBG FP resonator with the PT-SBG inside the cavity with the following spacings $\Lambda = 0.5$ µm; $L_1 = L_2 = 2000\Lambda = 1$ mm; $L_{PT} = 20000\Lambda = 10000$ µm, $d_1 = 100.5\Lambda = 50.25$ µm ; $d_2 = 1000.5\Lambda = 500.25$ µm. In the upper panels, with parameters $\kappa_1 = \kappa_2 = 1257$ m$^{-1}$ and $\kappa_{PT} = 202.63$ m$^{-1}$, the reflectivities of the left and right DBR are equal, while in the lower panels, with parameters $\kappa_1 = 0$; $\kappa_2 = 3770$ m$^{-1}$ and $\kappa_{PT} = 100$ m$^{-1}$, the reflectivities are 0 and $\approx 100\%$ respectively. The inserts in (a) and (c) show the reflection spectra of the left (red), right (blue) and PT-symmetric Bragg grating (magenta).

### 3.2.2 Suppression of left reflection

It is easy to see that when the PT-SBG strength is close to
$$(\kappa_{PT}L_{PT})_0^{(left)} = \coth(\kappa_1 L_1) + \coth(\kappa_2 L_2) \tag{25}$$
the FP structure becomes reflectionless from the left side, while the transmission and right-side reflection are given by:
$$\left|r_{PT}^{(right)}\right| = \frac{\cosh(\kappa_1 L_1 - \kappa_2 L_2)\sinh(\kappa_1 L_1 + \kappa_2 L_2)}{\sinh^2(\kappa_2 L_2)}; \qquad |t_{PT}| = \frac{\sinh(\kappa_1 L_1)}{\sinh(\kappa_2 L_2)} \tag{26}$$

Unlike the case of the fully constructive cavity interaction, the condition for zero reflection is fully symmetrical, i.e. exchanging the grating strengths, $\tanh(\kappa_1 L_1) \Leftrightarrow \tanh(\kappa_2 L_2)$, does not affect the zero reflectivity condition as long as $\tanh(\kappa_1 L_1)$ and $\tanh(\kappa_2 L_2)$ satisfy Eq.(25). However, the reflectivity of the left and right Bragg gratings drastically affects the transmission and right side reflectivity of the FP resonator. When the reflectivities of the left and right Bragg reflectors are the same, $\tanh(\kappa_1 L_1) = \tanh(\kappa_2 L_2) = 0.858$, the reflectivity for the left side is zero at the resonance wavelength (1550 nm), see Fig.6d, whereas the reflection from the right side occurs with strong (more than 5 times) amplification, see Fig.6f, with unamplified and non-attenuated transmission (Fig.6e). When the reflectivity of the left Bragg grating $\tanh(\kappa_1 L_1) = 0.996$ is increased and the right one is decreased, $\tanh(\kappa_2 L_2) = 0.736$, the transmission window inside the reflection spectrum from the left side becomes extremely narrow (Fig.6a), and the transmission and reflection from the right side are amplified (about 10 dB (Fig.6b) and 100 dB (Fig.6c), respectively).This situation is very similar to that shown in Figs.4a, 4b and 4c for the configuration with the fully constructive cavity coupling. However, when the reflectivities of the right and left Bragg gratings are exchanged ($\tanh(\kappa_2 L_2) = 0.996$; $\tanh(\kappa_1 L_1) = 0.736$), the FP resonator exhibits a very unusual behavior: the left- side reflectivity is still zero at the resonance wavelength (Fig.6g), i.e. input light launched from left side will pass into the cavity, but it will not pass through, because the transmission is very close to zero (Fig.6h). Finally, the right-side reflectivity is very close to the reflectivity of the right-side Bragg reflector (Fig.6i).

This spectral behavior with light launched from the left side is reminiscent of the so-called Coherent Perfect Absorber (CPA), the time-reversed counterpart of laser emission, when all the incoming coherent light can be trapped and converted to some form of internal energy [9, 10]. Even with a very strong distributed reflector, left or right, a small fraction of the light penetrates into the cavity with the PT-SBG inside. As an input light wave passes into the resonator it reflects back from the reflective end of the PT-SBG with strong enhancement and from the second distributed reflector, then the reflected back wave is reflected again by the first reflector. Through numerous iterations of such a process, the multiple wave replicas could interact constructively, producing a very narrow band-gap in the reflection and transmission spectra, as in Fig.6(a) and 6(b), or they could interact destructively and change the reflection and transmission very little, as in Fig.6(h) and 6(i).

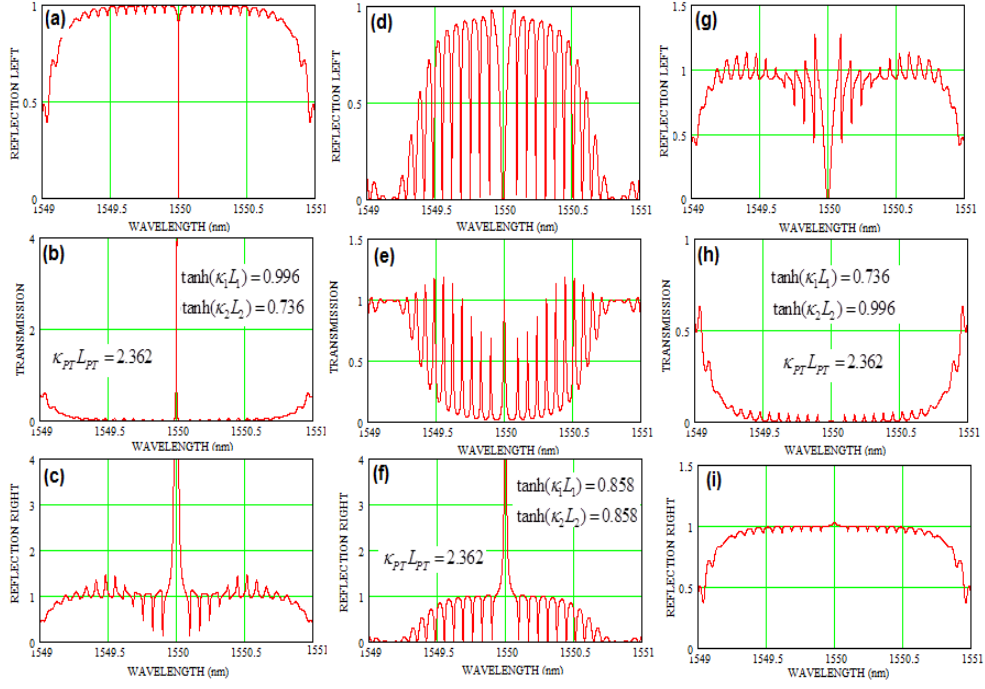

Fig.6. Partially constructive interference with left-side reflection suppressed at resonance: left-side reflection spectra (a, d, g), transmission spectra (b, e, h) and right-side reflection spectra (c, f, i) of the DBG FP resonator with the PT-SBG inside the cavity with the following structure parameters: $\Lambda = 0.5$ μm; $L_1 = L_2 = 2000\Lambda = 1$ mm; $L_{PT} = 20000\Lambda = 10000$ μm, $d_1 = 100.5\, \Lambda = 50.25$ μm ; $d_2 = 1000.5\Lambda = 500.25$ μm; (a), (b) and (c) $\kappa_1 = 3142$ m$^{-1}$; $\kappa_2 = 942.5$ m$^{-1}$; $\kappa_{PT} = 236.18$ m$^{-1}$; (d), (e) and (f) $\kappa_1 = \kappa_2 = 1287$ m$^{-1}$; $\kappa_{PT} = 236.18$ m$^{-1}$; (g), (h) and (i) $\kappa_1 = 942.5$ m$^{-1}$; $\kappa_2 = 3142$ m$^{-1}$; $\kappa_{PT} = 236.18$ m$^{-1}$.

### 3.2.3 Suppression of right reflection

We can also set the PT-SBG strength to suppress reflection from the right side. It is obvious from Eq.(24) that the following relationship between the gratings should be maintained:

$$(\kappa_{PT} L_{PT})_0^{(right)} = \tanh(\kappa_1 L_1) + \tanh(\kappa_2 L_2) \tag{27}$$

Then the transmission and left-side reflection take the following form:

$$\left| r_{PT}^{(left)} \right| = \frac{\cosh(\kappa_1 L_1 - \kappa_2 L_2)\sinh(\kappa_1 L_1 + \kappa_2 L_2)}{\cosh^2(\kappa_1 L_1)}; \qquad \left| t_{PT} \right| = \frac{\cosh(\kappa_2 L_2)}{\cosh(\kappa_1 L_1)}; \tag{28}$$

The left-side, right-side reflection and transmission spectra are shown in Fig.7 for different reflectivities of the left and right Bragg reflectors ( $\tanh(\kappa_1 L_1)$ and $\tanh(\kappa_2 L_2)$ ). Similarly to the zero left-side reflectivity case, the condition for zero right-side reflectivity is symmetrical with respect to $\tanh(\kappa_1 L_1)$ and $\tanh(\kappa_2 L_2)$, as follows from Eq.(26). However, a different ratio of $\tanh(\kappa_1 L_1)$ to $\tanh(\kappa_2 L_2)$ produces very different results for the transmission and left-side reflection of the FP-resonator. For left and right reflectors of identical strength, $\tanh(\kappa_1 L_1)/\tanh(\kappa_2 L_2) = 1$, the transmission at the resonance wavelength is unity, with two sharp amplifying peaks from left and right (barely visible in Fig.7e). The left-side reflection occurs with strong amplification. The right-side reflection is zero at the resonance, with two sharp amplifying peaks from left and right (Fig.7f). For $\tanh(\kappa_1 L_1)/\tanh(\kappa_2 L_2) < 1$ the situation is very similar to that shown in Fig.6(g), 6(h) and 6(i), only for input light launched from right side.

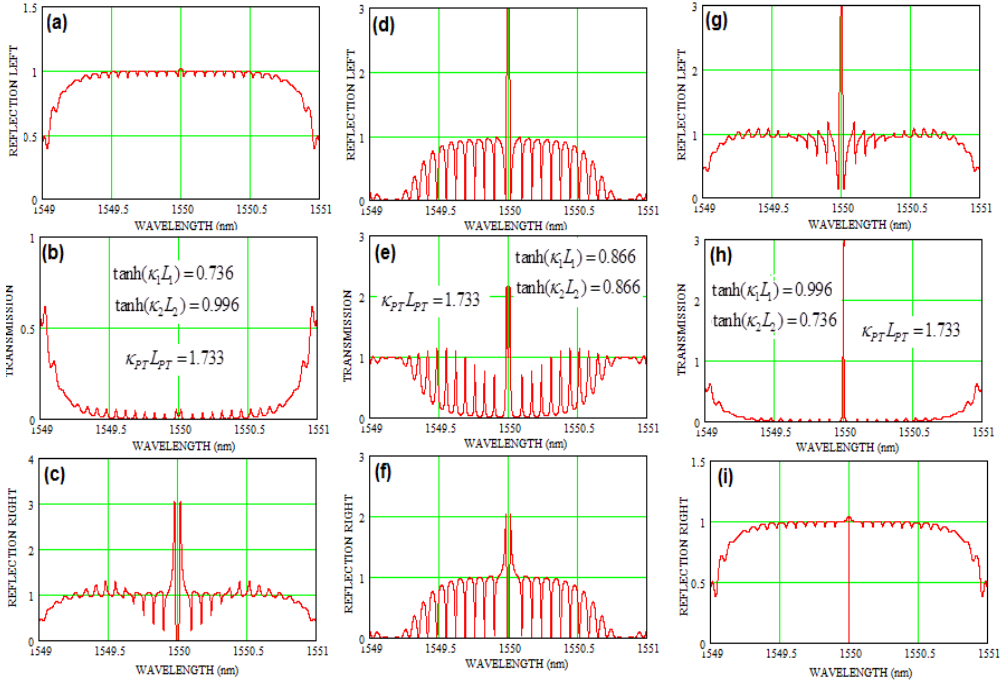

Fig.7 Partially constructive interference with left-side reflection suppressed at resonance: left-side reflection spectra (a, d, g), transmission spectra (b, e, h) and right-side reflection spectra (c, f, i) of the DBG FP resonator with the PT-SBG inside the cavity with the following structure parameters: $\Lambda = 0.5$ μm; $L_1 = L_2 = 2000\Lambda = 1$ mm; $L_{PT} = 20000\Lambda = 10000$ μm, $d_1 = 100.5\Lambda = 50.25$ μm ; $d_2 = 1000.5\Lambda = 500$ μm; (a), (b) and (c) $\kappa_1 = 3142$ m$^{-1}$; $\kappa_2 = 942.5$ m$^{-1}$; $\kappa_{PT} = 173.26$ m$^{-1}$; (d), (e) and (f) $\kappa_1 = \kappa_2 = 1287$ m$^{-1}$; $\kappa_{PT} = 173.26$ m$^{-1}$; (g), (h) and (i) $\kappa_1 = 942.5$ m$^{-1}$; $\kappa_2 = 3142$ m$^{-1}$; $\kappa_{PT} = 173.26$ m$^{-1}$.

### 3.3. Partially destructive cavity interaction

When condition (16c) is fulfilled, the left-side and right-side reflectivity and the transmission have the following expressions at the resonance ($\sigma=0$):

$$\left| r_{PT}^{(left)} \right| = \left| \frac{\tanh(\kappa_1 L_1) + \kappa_{PT} L_{PT} \tanh(\kappa_1 L_1)\tanh(\kappa_2 L_2) - \tanh(\kappa_2 L_2)}{1 + \kappa_{PT} L_{PT} \tanh(\kappa_2 L_2) - \tanh(\kappa_1 L_1)\tanh(\kappa_2 L_2)} \right|;$$

$$\left| r_{PT}^{(right)} \right| = \left| \frac{\tanh(\kappa_2 L_2) + \kappa_{PT} L_{PT} - \tanh(\kappa_1 L_1)}{1 + \kappa_{PT} L_{PT} \tanh(\kappa_2 L_2) - \tanh(\kappa_1 L_1)\tanh(\kappa_2 L_2)} \right|; \tag{29}$$

$$\left| t_{PT} \right| = \frac{\text{sech}(\kappa_1 L_1)\text{sech}(\kappa_2 L_2)}{\left| 1 + \kappa_{PT} L_{PT} \tanh(\kappa_2 L_2) - \tanh(\kappa_1 L_1)\tanh(\kappa_2 L_2) \right|}.$$

#### 3.3.1 Lasing

The DBR in such a configuration is not able to lase because the threshold condition reduces to the following equation:

$$\kappa_{PT} L_{PT} = \tanh(\kappa_2 L_2) - \coth(\kappa_1 L_1), \tag{30}$$

which cannot be satisfied since the grating strength must be positive.

### 3.3.2 Suppression of left reflection

However, the strength of the PT-SG ($\kappa_{PT} L_{PT}$) can be set to suppress reflectivity from the left side to zero when

$$(\kappa_{PT} L_{PT})_0^{(left)} = \coth(\kappa_1 L_1) - \coth(\kappa_2 L_2), \tag{31}$$

which requires $\kappa_2 L_2 > \kappa_1 L_1$. The right-side reflection and the transmission then reduce to the following equations:

$$\left| r_{PT}^{(right)} \right| = \frac{\cosh(\kappa_1 L_1 + \kappa_2 L_2) \sinh(\kappa_2 L_2 - \kappa_1 L_1)}{\sinh^2(\kappa_2 L_2)}; \qquad \left| t_{PT} \right| = \frac{\sinh(\kappa_1 L_1)}{\sinh(\kappa_2 L_2)}. \tag{32}$$

### 3.3.3 Suppression of left reflection

The PT-SG strength can also be adjusted to fully suppress reflection from the right side:

$$(\kappa_{PT} L_{PT})_0^{(right)} = \tanh(\kappa_1 L_1) - \tanh(\kappa_2 L_2), \tag{33}$$

which requires $\kappa_1 L_1 > \kappa_2 L_2$. The left-side reflection and transmission are then as follows:

$$\left| r_{PT}^{(left)} \right| = \frac{\cosh(\kappa_1 L_1 + \kappa_2 L_2) \sinh(\kappa_1 L_1 - \kappa_2 L_2)}{\cosh^2(\kappa_2 L_2)}; \qquad \left| t_{PT} \right| = \frac{\cosh(\kappa_1 L_1)}{\cosh(\kappa_2 L_2)}. \tag{34}$$

The right-side and left-side reflections and transmission for the last case of partially destructive cavity interaction are summarized in Fig.8. In this configuration a behavior very similar to Fig.6 and 7 is achieved under very low PTS_BG strength (below its amplification level).

### 3.4. Fully destructive cavity interaction

Finally the condition (16d) is responsible for fully destructive cavity interaction, when there is no physical value of the PT-SBG strength that can provide lasing, with the following expressions for reflections and transmission:

$$\left| r_{PT}^{(left)} \right| = \frac{\tanh(\kappa_2 L_2) + \kappa_{PT} L_{PT} \tanh(\kappa_1 L_1) \tanh(\kappa_2 L_2) + \tanh(\kappa_1 L_1)}{1 + \kappa_{PT} L_{PT} \tanh(\kappa_1 L_1) + \tanh(\kappa_1 L_1) \tanh(\kappa_2 L_2)}$$

$$\left| r_{PT}^{(right)} \right| = \frac{\tanh(\kappa_1 L_1) + \kappa_{PT} L_{PT} + \tanh(\kappa_2 L_2)}{1 + \kappa_{PT} L_{PT} \tanh(\kappa_1 L_1) + \tanh(\kappa_1 L_1) \tanh(\kappa_2 L_2)} \tag{35}$$

$$\left| t_{PT} \right| = \frac{\operatorname{sech}(\kappa_1 L_1) \operatorname{sech}(\kappa_2 L_2)}{1 + \kappa_{PT} L_{PT} \tanh(\kappa_1 L_1) + \tanh(\kappa_1 L_1) \tanh(\kappa_2 L_2)}$$

Such a configuration does not produce any interesting optical properties.

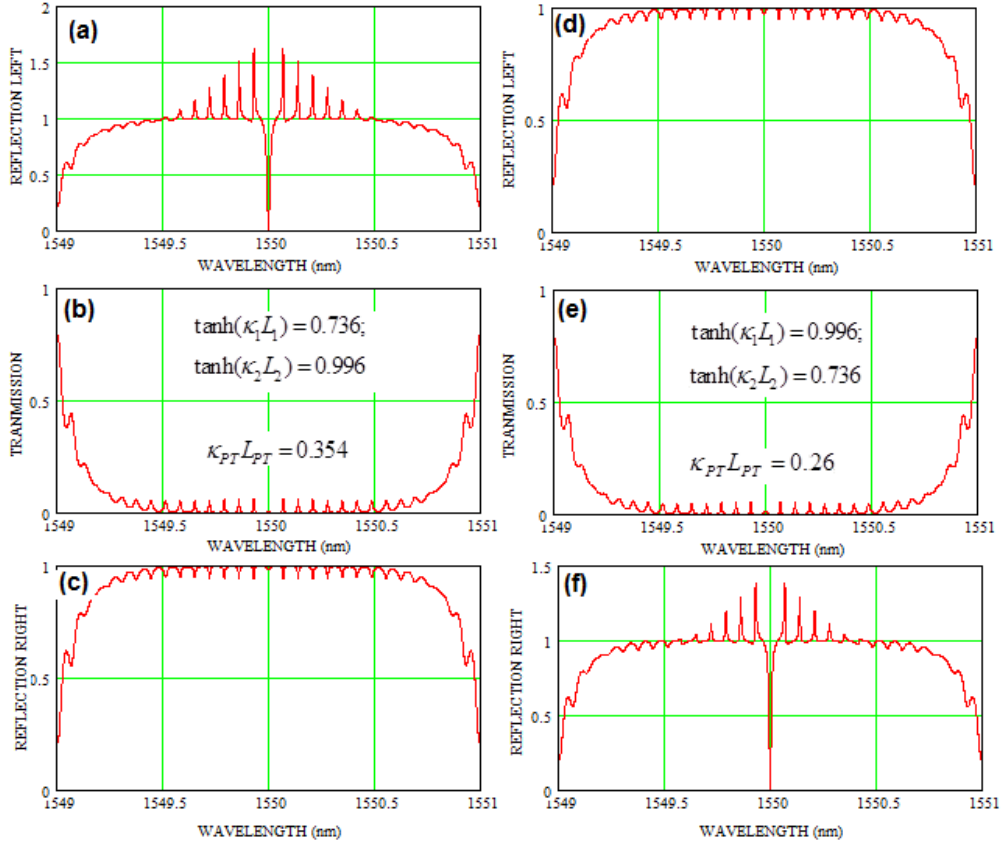

Fig.8. Partially destructive interference: Left-side (a) and (d), right-side (c) and (f) reflection and transmission (b) and (e) spectra of the DBG FP resonator with the PT-SBG grating inside the cavity. The spacings $\Lambda$ = 0.5 µm; $L_1 = L_2 = 2000\Lambda$ = 1 mm; $L_{PT} = 20000\Lambda = 10000$ µm, $d_1$ = 100.5 $\Lambda$ =50.25 µm ; $d_2$ =1000$\Lambda$=*500* µm are common to all panels. In the left panels, with parameters $\kappa_1$ = 942.48 m$^{-1}$; $\kappa_2$=3142 m$^{-1}$; $\kappa_{PT}$ =35.42 m$^{-1}$, left-side reflection is suppressed at resonance, while in the right panels, with parameters $\kappa_1$ = 3142 m$^{-1}$; $\kappa_2$=942.48 m$^{-1}$; $\kappa_{PT}$ =32.61 m$^{-1}$, right-side reflection is suppressed at resonance.

## 4. Conclusion

A composite cavity consisting of a Fabry-Perot resonator with distributed Bragg reflectors and with an embedded PT-symmetrical Bragg grating was proposed and theoretically investigated. This PT-symmetrical Bragg grating at its breaking point is completely invisible for a light wave travelling in one direction, and strongly reflective for the wave propagating in the opposite direction. The unique asymmetrical reflectivity of the PTS-BG inside a traditional Fabry-Perot resonator basically produces two independent embedded cavities. Depending on the resonator geometry and mutual arrangement of these cavities, they can interact with different degrees of coherency: fully constructive interaction, partially constructive interaction, partial distractive interaction, and finally their interaction can be completely destructive. We analyze the transmission and reflection properties of these new structures through a transfer matrix approach. A number of very unusual (exotic) nonsymmetrical absorption and amplification behaviors are observed. For example, the presence of the PTS-BG can produce a narrow transmission band in a practically 100% distributed Bragg reflector. So input light penetrates into the resonator, where it fully absorbed. The same input signal launched from the opposite side of the resonator will be fully reflected. In an alternative arrangement the resonator passes the signal through the resonator within a very narrow bandwidth around the resonance wavelength when it is launched from one side; however, it is reflected with significant amplification when launched from the opposite end.

The proposed structure also exhibits unusual lasing characteristics. The presence of the PTS-BG with very weak strength (very small index and loss/gain modulation) forces the resonator to generate a single mode irrespective of the total resonator length. Such lasing occurs with zero level of gain in the resonator. Due to the PT-symmetrical grating, there is no chance of mode hopping; it can lase with only a single longitudinal mode for any distance between the distributed reflectors. The proposed structures pave a novel route for designing a new class of multifunctional optical devices with completely nonsymmetrical optical responses and new mechanisms for light manipulation.

**References**

1. Li Ge, Y. D. Chong, S. Rotter, H. E. Türeci, and A. D. Stone, "Unconventional modes in lasers with spatially varying gain and loss," Physical Review *A* **84**, 023820 (2011).
2. H. E. Türeci, A. D. Stone, and B. Collier, "Self-consistent multimode lasing theory for complex or random lasing media," Phys. Rev. A **74**, 043822 (2006).
3. C.M. Bender, S. Boettcher, P.N. Meisinger, "PT -Symmetric Quantum Mechanics," J. Math. Phys. **40**, 2201 (1999).
4. L. Poladian, "Resonance mode expansions and exact solutions for nonuniform gratings," Phys. Rev **54** (3), 2963–2975 (1996).
5. H.-P. Nolting, G. Sztefka, M. Grawert, and J. Ctyroky, "Wave Propagation in a Waveguide with a balance of Gain and Loss," in Integrated Photonics Research' 96, (OSA, 1996) 76-79.
6. M. Kulishov, J.M. Laniel, N. Belanger, J. Azana, and D.V. Plant, "Nonreciprocal waveguide Bragg gratings," Opt. Express. **13**, 3068-3078 (2005).
7. M. Greenberg, M. Orenstein, "Irreversible coupling by use of dissipative optics," Opt. Lett., **29**, 451-453 (2004).
8. M. Kulishov, B. Kress, R. Slavik, "Resonant cavities based on Parity-Time symmetric diffractive gratings," Opt. Express, **21**, 9473-9483, (2013).
9. C.Y Huang, R. Zhang, J.L. Han, J. Zheng, and J.Q. Xu, "Type II perfect absorption modes with controllable bandwidth in PT-symmetric/Traditional Bragg grating combined structures," Phys. Rev **A**89, 023842 2014.
10. S. Longhi, "PT-symmetric Laser-absorber," Phys. Rev. A **82**, 031801 (2010).
11. Li Ge, Y.D. Chong, and A.D. Stone, "Conservation relations and anisotropic transmission resonances in one dimensional PT-symmetric photonic heterostructures," Phys. Rev. A **85**, 023802 (2012).
12. Y. Zhao, J. Chang, Q. Wang, J. Ni, Z. Song, H. Qi, C. Wang, P. Wang, L. Gao, Z. Sun, G. Lv, T. Liu, G. Peng, "Research on novel composite structure $Er^{3+}$ doped DBR fiber l laser with a π-phase shifted FBG," Opt. Express. **21**, 22515-22522 (2013).